\documentclass[]{svmult}

\usepackage{mathptmx}       
\usepackage{helvet}         
\usepackage{courier}        
\usepackage{type1cm}        
\usepackage{makeidx}         
\usepackage{graphicx}        
\usepackage{multicol}        
\usepackage[bottom]{footmisc}
\usepackage{url}
\usepackage{cite}

\makeindex            

\begin{document}

\title{Morphological transformations  of Dwarf Galaxies in the Local Group}
\titlerunning{Morphological transformation of Dwarf Galaxies} 
\author{{\bf Giovanni Carraro}}
\authorrunning{G. Carraro} 
\institute{Giovanni Carraro \at European Southern Observatory, Alonso de Cordova 3107, 19001, Santiago de Chile, Chile, \email{gcarraro@eso.org}}

\maketitle

\abstract{ In the Local Group there are three main types of dwarf galaxies: Dwarf Irregulars, Dwarf Spheroidals, and Dwarf Ellipticals.
Intermediate/transitional types are present as well.
This contribution reviews the idea that the present day variety of  dwarf galaxy morphologies in the Local Group might reveal the existence
of a transformation chain of events, of which any particular  dwarf galaxy  represents a manifestation of a particular stage.
In other words,  all dwarf galaxies  that now are part of the Local Group
would have formed identically in the early universe, but then evolved differently because of morphological transformations induced 
by dynamical processes like galaxy harassment, ram pressure stripping,  photo-evaporation, and so forth.
We start describing the population of dwarf galaxies and their spatial distribution in the LG.  Then, we describe those phenomena that can alter the morphology of a dwarf galaxies, essentially by  removing, partially or completely, their gas content. Lastly,
we  discuss morphological signatures in the Local Group Dwarf Galaxies that can be attributed to different dynamical phenomena.
While it is difficult to identify  a unique and continuous transformation sequence, we have now a reasonable understanding of the basic
evolutionary paths that lead to the various dwarf galaxy types.}

\section{Introduction}
\label{sec:1}
The Local Group (LG)  has a physical radius of $\sim$1.2 Mpc (Van den Bergh 1999): this 
is defined as the radius  of its zero-velocity  surface, namely the surface which separates the LG from the  field expanding with the Hubble flow. The LG is located at the outskirts of the large Virgo cluster.
The actual LG  is dominated by three spiral galaxies:  M~31(NGC~224, the Andromeda galaxy) , the Milky Way (MW), and M~33 (NGC~598, the Triangulum galaxy), in decreasing order of mass. 
There are no giant or intermediate-mass ellipticals in the LG.  
The remaining galaxies are dwarf galaxies (DG), and their number has been increasing over the years, since fainter and fainter objects are being discovered (e.g. Crater, Belokurov et al. 2014) out of deep and wide area  sky surveys (2MASS, SDSS, etc.). The exact number of DG in the LG is unknown.  As of today,  they would be $\sim$ 70 (McConnachie 2012).\\

The majority of these DGs groups around M~31 and the MW.
M~33, the smallest galaxy among the spirals,  might have two possible companions (the Pisces and Andromeda XXII dwarfs).
All the remaining galaxies are too distant from the 3 dominant spirals to be considered bound to them.  
NGC~3109 and Antlia may constitute a group themselves (the 14$+`$12 group, McConnachie 2012) , together with Sextans A and B, and, possibly , Leo P (Bernstein-Cooper et al. 2014).   The membership of the NGC~3109 - and companions- to the LG is  disputed, since this group shows a filamentary structure and seems to be in the verge to enter the LG (Bellazzini et al. 2013, 2014) for the first time.\\

In Figs.~1 and 2 we show the distribution of DG in the LG according to the recent compilation by McConnachie (2012). Fig.~ 1 shows the DGs  associated with the Milky Way, while Fig.~2  shows  DGs associated with M~31 (blue symbols), and DGs  that are believed not to be associated with the two major spirals (green symbols). 
A few DG discovered after 2012 (like Crater) are missing in this plot.
Let me also mention that  Segue 1 is still a controversial object (it might be globular cluster, Niederste-Ostholt et al. 2009), and, lastly,  that Canis Major has been ruled out as a DG (Momany et al 2006).

\begin{figure}[]
\includegraphics[scale=.5,angle=-90]{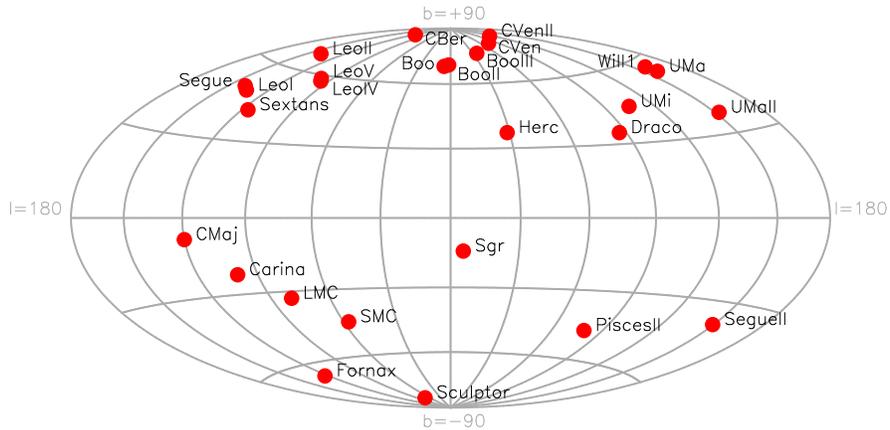}
\caption{Distribution of dwarf galaxies in the LG following McConnachie 2012: DGs associated with the MW (red symbols).}
\label{fig:1}       
\end{figure}

\begin{figure}[]
\includegraphics[scale=.5, angle=-90]{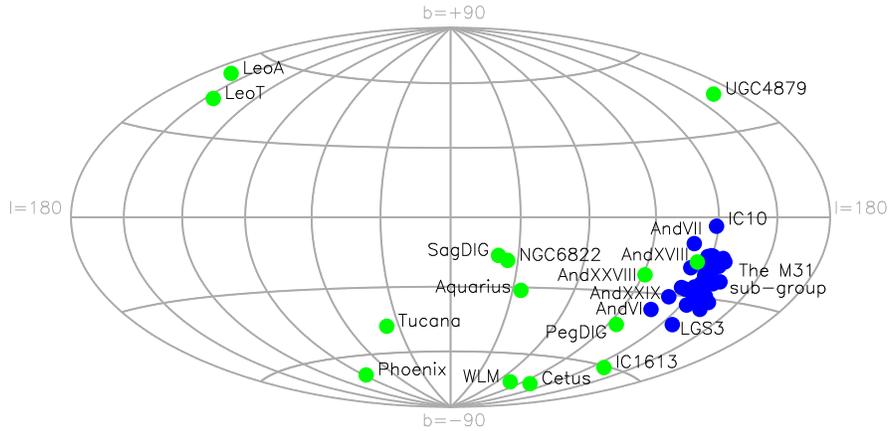}
\caption{Distribution of dwarf galaxies in the LG following McConnachie 2012:DGs associated with M~31 (blue symbols), and DGs considered not associated with these 2 major spirals.}
\label{fig:2}       
\end{figure}

Many exhaustive reviews have been written on the LG  structure (Grebel 1997; Mateo 1998, van den Bergh 2000), and  
I refer the reader to them for all additional details.

Instead, in this contribution I shall focus on the morphology of DGs, and to the variety of structures they exhibit. In many occasions
this variety has been seen as an evolutionary chain (Moore et al. 1996), in which more dynamical evolved galaxies contain less and less gas.  

\begin{table*}
\caption{Local Group dwarf galaxies basic physical parameters}
\begin{tabular}{lccc}
\hline
Morphological type & $\mu_V$ & $M_{HI}$ & $M_{tot}$\\
\hline
   & $mag \times arcsec^{-2}$ & $M_{\odot} $& $M_{\odot}$\\
\hline\hline
Dwarf Ellipticals      &   $\leq 21$ &  $\leq 10^8$& $\leq 10^9$ \\
Dwarf Spheroidals &   $\geq 22$&  $\leq 10^5$& $\sim 10^7$ \\
Dwarf Irregulars      &  $\leq 23$  &  $\leq 10^9$ &  $\leq 10^{10}$  \\
\hline\hline
\end{tabular}
\end{table*}

\section{Taxonomy of Dwarf Galaxies in the Local Group}
\label{sec:2}
Dwarf galaxies in the LG divide into 3 main morphological types: Spheroidals ({\bf dSph}), Ellipticals ({\bf dE}) , and Irregulars ({\bf dIrr}). 
Intermediate type are also known to exist in the LG.\\
Using Table~1 as a guideline, we can briefly describe their properties as follows:

\begin{description}

\item $\bullet$ {\bf Dwarf Irregulars}:  these are atomic-gas (HI) dominated systems, and exhibit a variety of  irregular shapes; SagDIG  (Young \& Lo 1997) and Sextans A (van Dyk et al. 1998) are two classical examples.  This type of DGs is found in any environment:  galaxy clusters, small galaxy groups, and in the general field. Several irregulars show a disk structure, but the disk does not seem to be ubiquitous.  Zhang et al. (2012) demonstrated that most dIrrs start their life with a disk that then shrinks (outside-in scenario), so it is conceivable that all dIrrs form as low-mass spiral galaxies with extended gas discs. 
The evolution of star formation (SF) and disk structure is, however, different from  large spirals, that are believed to assemble via an inside out scenario.
Over their lifetime isolated dIrrs  kept an almost constant SF (e.g. IC 1613, SagDIG), although some of them show signatures of a pristine significant peak of star formation (Skillman et al. 2014, Momany et al 2005), in close similarity with Blue Compact Dwarfs.  Whenever searched for, old stellar populations have been detected in irregular galaxies (see Leo I, Held et al. 2001). dIrr tend to be isolated systems.\\

\item $\bullet$ {\bf Dwarf Ellipticals}:  these DGs have a more regular shape, but are not depleted of gas. In  the LG all the known  dEs are located around M 31. The most known case is the one of NGC 205 (Monaco et al. 2009).  They host a mixture of stellar populations, and show a complex star formation history (Carraro et al. 2001). They contains old, intermediate age and young stellar population.
As for the dynamics, they are rotation supported. These galaxies tend to be concentrated closer to M31.\\

\item $\bullet$ {\bf Dwarf Spheroidals}: These are dynamically evolved stellar systems, composed of intermediate age to old stellar populations.  In most cases they are random motion supported, like Elliptical galaxies, although mild rotation has been detected in some of them.  Being loose and dispersed systems, they are believed to be DM dominated. This is however based on the assumption that these systems are in virial equilibrium, which is not really completely proved (Lughausen et al.  2014).
Typically,  dSph are found close to M 31 or the MW. They are almost  devoid of  gas in  their central
   parts , although gas has been detected in the surrounding of some of them (Sculptor and Phoenix, for example). \\

\end{description}

A few dwarfs such as Phoenix, Pegasus, Antlia, DDO 210, and LGS 3, are classified as transition
objects (dIrr/dSph); possibly evolving from dIrr to dSph (Grebel et al. 2003). The idea of a single evolutionary endpoint
is rather appealing: dE would be the remnants of dIrr that have lost their gas (by stripping in clusters
or near large galaxies, i.e. an environmental effect), while dIrr/dSph would testify to an intermediate
case of dwarfs that have maintained some of their gaseous content.\\

\noindent
We believe there are no Ultra Compact Dwarf ({\bf UCDs}) in the LG, although this depends on the precise definition of UDC 
in terms of mass and luminosity (Mieske et al. 2012). It might be that M~32 or Omega Cen are  UCDs. There is some consensus that UCD are not precisely DGs, but simply the high-mass tail of the globular clusters mass function (Mieske et al. 2012) .
Lastly, the LG does not contain any Blue Compact or Starburst Blue Compact Dwarf ({\bf BCD,SBCD}).\\

\begin{figure*}[]
\includegraphics[scale=.5, angle=-90.]{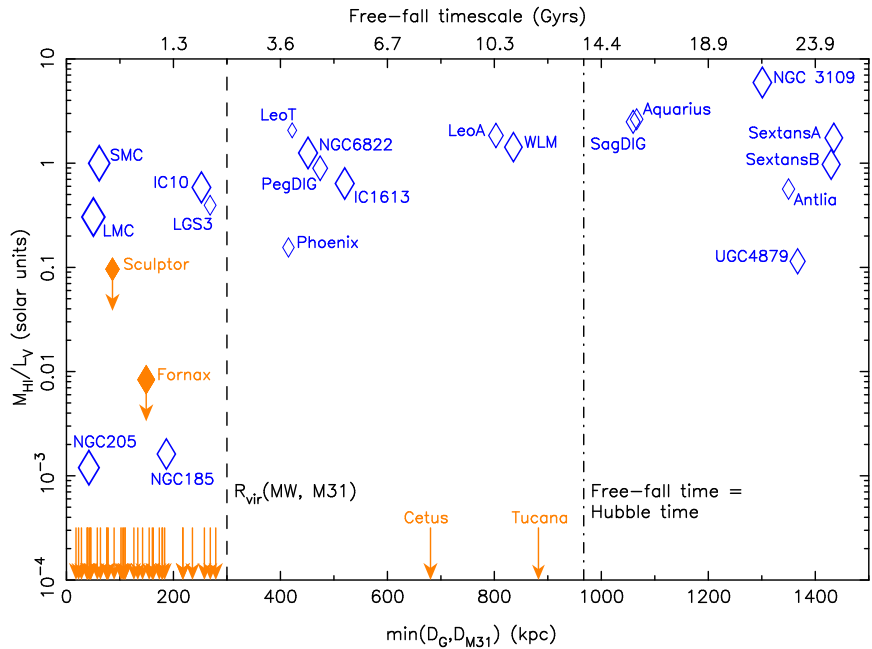}
\caption{Morphological segregation of DGs in the LG from McConnachie (2012).}
\label{fig:3}       
\end{figure*}

\section{Spatial distribution of different dwarf galaxy types in the Local Group}
\label{sec:3}
The first indication of a possible morphological evolution of DGs comes from their spatial distribution.
Different morphological type are not randomly  distributed  across the LG volume, but shows a marked morphological segregation.\\
Gas poor DGs (dE and dSph) cluster close to  MW and M31, while gas-rich DGs (dIrr) tend to be spread over a much larger volume (see Fig.~3). 
This is only on the average, since there are several exceptions. \\
dSph concentrates around M 31 and and the MW, while dEs are found
in the surroundings of M31 only. \\ Cetus and Tucana are two examples of dSph not closely located to major spirals (see Fig.~3).\\

\section{ Morphological transformations}
\label{sec:4}
Structural modifications can occur in DGs as a result of  a few dynamical (internal or environmental) processes:

\subsection{Ram Pressure stripping}
Ram pressure stripping occurs when a DG is moving within a sufficiently dense multi-phase intra-cluster (ICM)  medium.
If the ICM ram pressure exceeds  its gravitational force (Mori \& Burkert  2000), a DG can loose partially or entirely its gas content.
Lin \& Faber (1983) firstly estimated that the ICM in the LG should have a density of the order of $10^{-6} cm^{-3}$ 
for ram pressure to be effective. Observational evidences are, however, scanty. The intra-group medium (IGM) in the LG seems to be multi-phase.\\
So far, observations allow us to identify two phases:  

\begin{description}
\item $\bullet$ cold  (T $\leq 10^{4} K$) gas has been detected in HI (Richter 2012), and tends to be concentrated in the periphery (within 50 kpc) of the two major spirals (Andromeda and the Milky Way);
\item $\bullet$ hot  (T $\geq 10^{4} K$) gas  has been detected via O VI emission lines in the UV wave-length regime (Sembach et al. 2003), and in X-Ray, via O VIII emission lines (Gupta et al. 2012). This hot gas tends to occupy the outer regions of the LG.
\end{description}

No clear figures are available for the density of these two components in the LG.
According to hydro-dynamical simulations (Nuza et al. 2014), the spatial distribution of these two components might be the result of the various processes which affects gas circulation, like inflows of extra-LG material and outflows from DGs internal or tidal evolution.\\

Nbody/gas-dynamical simulations (Mayer et al. 2006) confirm that ram-pressure can be the most efficient process to remove gas from a DG
in a cluster of galaxies, but that this depends a lot on the evolutionary status of the DG and on the ICM (see also Steinmetz, this conference). 

\subsection{Harassment}
Most DGs are believed to be orbiting around one of the major spirals in the LG.  NGC~3109 and Antlia seem to constitute
a separate group (the 14$+$12 group, Bellazzini et al. 2013, Bernstein-Cooper et al. 2014).
There are also a number of DGs that are too distant to be bound to either the MW or Andromeda.
With the term {\it harassment} we refer to the tidal interaction exerted by a major galaxy on an orbiting DG. The result
is typically tidal stirring of the DG which, in extreme cases, can also lead to tidal stripping and removal of the gas.\\

There is nowadays a lot of work in the field of dwarf galaxies orbits. This is a major requirement if one wants to understand 
their dynamical evolution and possible origin.
The best known case is the LMC/SMC pair: accurate proper motions have been provided in many occasions, and modern orbits calculated with the aim to understand their complex 3 body problem. However, out of the very same figures
different scenarios for the past orbital evolution, and the implied SFH and origin of the clouds have been discussed
(Kallivayalil et al. 2013, Besla, this conference) and still rely on strong assumptions on the 3D structure of the clouds and their possible binary nature. In any case
the existence of the Magellanic bridge and  stream clearly  indicates that tidal interaction and induced SF is ongoing (Casetti-Dinescu et al 201; see also Fukui and Gallart, this conference).

The other obvious case is the one of the Sgr dSph (Ibata et al. 1994, Majewski, this conference), whose orbit is better constrained. The tidal arms are traced all over the Galactic halo (Majewski et al. 2004) , and the Age-Metallicty relation (AMR, a consequence of the SFH) well defined.\\

As for the other MW dwarf galaxies,  proper motions have been estimated for Fornax, Leo I, Sculptor, Draco, Carina, and Ursa Minor, and attempts have been done to reconstruct their orbits (Pasetto et al. 2003, 2009)
and spatial configuration (Pawlowski \& Kroupa 2013; Kroupa, this conference) . However, uncertainties
associated to proper motion measurements are still too large to derive solid orbital solutions.

For M31 satellites, only 2 dwarf galaxies have measured proper motions (Watkins 2013), and for all the others proper motions expectations have been derived from dynamical considerations only.

\subsection{Internal stellar evolution}
One possibility for a DG to loose its gas content is via stellar evolution. As a consequence of a strong burst of star formation, SNae and stellar winds inject kinetic energy in the surrounding gas. If the gas acquires a velocity high enough to over-pass the galaxy potential
well, it can escape the galaxy (blown-away) after having been displaced into the halo (blown-out) from the disk. This phenomenon has been analytically and numerically studied by McLow and Ferrara (1999) as a function
of the DM halo mass.  In order for this to happen, a strong collimated  burst of star formation needs to occur, which is not seen in LG DGs
(Recchi \& Hensler 2013) . It might be possible that BCD experiences such strong bursts of star formation, but there are no such DGs in the LG. Evidences of super bubbles which can be indicative of gas escaping from a DG have been found outside the LG (in I Zw 18 and NGC 1705).

\subsection{Photo-evaporation}
This can occur early during DG evolution in presence of strong UV radiation generated by the cosmic re-ionisation (Barkana \& Loeb 1999).
All galaxies  formed before  re-ionization having velocity dispersion lower than $\sim$ 10 km/sec can not survive, since they are
 turned into completely dark galaxies after the gas evaporation. When we analyse the Star Formation History in LG DGs, however,
 we do not find any indication of pristine abrupt interruption (quenching) of SF, which would be the case if gas is suddenly removed.  This implies that re-ionisation did not have any impact in the SFH of DG, but that other processes conspired to shaped it (Grebel \& Gallagher  2004; Hidalgo et al 2013, Steinmetz, this conference ).

\section{Examples in the Local Group}
\label{sec:5}
Having illustrated the physical processes that we believe are responsible for  the gas removal from dwarf galaxies, we 
will now look more closely at the Local Group, and search for any signature of these phenomena among its population of dwarf galaxies.
As we already mentioned, LG DG are divided in three main type: dSph, dE,  and dIrr. They differ in the amount of gas and in their structural, dynamical, and stellar evolution properties. Intermediate types are also present, mainly transitional dSph/dIrr (Grebel 1999). These latter are
of paramount importance since they help us to delineate a possible evolutionary path which  transforms a dIrr into a dSph via one or more of the processes described above. Transitional dSph/dIrr have been suggested as the possible progenitor of classical dSph.

\subsection{Pegasus: ram stripping caught in the act}
Optical and HI observations as described in  McConnachie et al. (2007) in the Pegasus isolated -but possibly associated with M3- dwarf have shown that stars and gas are distributed in a different way (see Fig.~4). While stars distribute in a disk-like structure, cold gas  exhibits a cometary shape. This is interpreted as Pegasus is moving across the intra-group medium and its gas is undergoing ram pressure stripping.  This would represent the best case of  ram pressure stripping in the LG, and would suggest the existence of significant hot gas in it. 
We remind that such gas has been recently detected around the Milky Way and Andromeda (Lehner et al. 2014), but not in the most isolated regions of the LG. 
Therefore Pegasus would represent an intermediate dwarf, say a dIrr  on the verge of turning - possibly- into a dE.

\begin{figure*}[]
\includegraphics[scale=.5, angle=-90.]{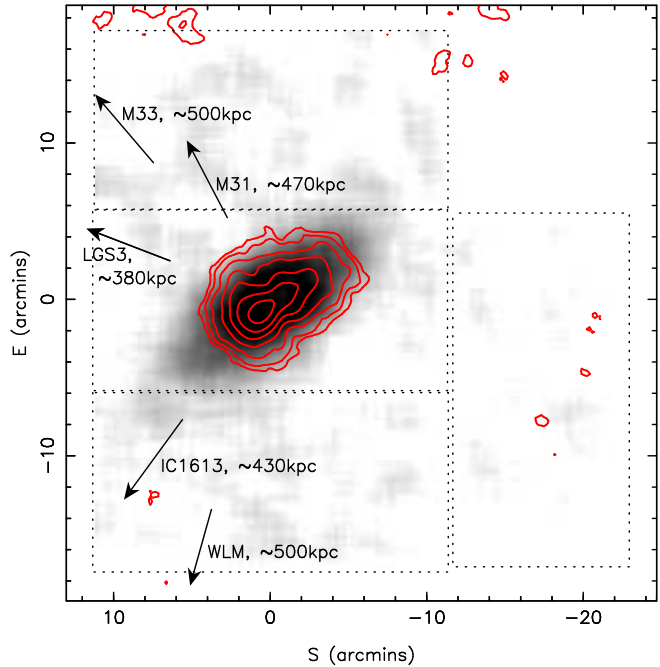}
\caption{The smooth disk-like star distribution in Pegasus, with over-imposed HI contours from McConnachie et al. (2007).}
\label{pig}       
\end{figure*}

\subsection{The 14+12 group: harassment at work}
Besides the prominent case of the Sgr dSph, there are numerous example of tidal interaction in the LG (e.g. the magellanic stream and bridge).
Bellazzini et al. (2013, 2014) draw the attention of NGC~3109 and its companion, and nicely showed that this group of 5 dwarfs 
(the 14+12 group: NGC~3109, Antlia, Sextans A, Sextans B, and Leo P) is moving along a filament. The observational evidences support  the conclusion that this group has been tidally disturbed ({\it harassed}), or that has been accreted as a filamentary sub-structure (a tidal tail) produced by some major interaction event. 
Bernstein-Cooper (2014) HI observations, however,  seem to rule out the membership of Leo P to this  group. Leo P, instead, would represent
the case for an isolated, gas rich, extremely metal deficient, dIrr.

\subsection{Leo I : extreme harassment}
The dSph galaxy Leo I possesses an extreme high radial velocity, and it seems to be unbound to both M~31 and the MW.  This dwarf is almost tidally disrupted, possibly as a consequence of several peri-galactic passages in its motion around  a massive galaxy (
either MW or M~31).
Unless its tangential motion  has been severely underestimated, this galaxy is on the verge of leaving the LG, totally devoid of gas.

\subsection{ Andromeda XII: untouched by the LG?}
An interesting, opposite case, has been reported a few years ago. This dSph galaxy seems to be infalling at very high speed into the LG (Chapman et al. 2007). Chapman et al. (2007) argue that, because of the large velocity,
this galaxy is entering the LG for the very first time, and therefore represents the best example of a late infall in the LG. Being depleted of gas,
this dSph must have experience strong tidal interactions before joining the LG.

\noindent
These examples, together with the Magellanic system and the Srg dSph, illustrate convincingly that DG in the LG are undergoing profound 
transformations.  Detailed HI observations, in tandem with ultraviolet spectroscopy of the medium surrounding each DG (Lehner et al. 2014) can help us to understand better their individual dynamical status and measure the relative important of any of the most important dynamical processes that occur in the LG environment. 

\section{Conclusions}
Due to their proximity, dwarf galaxies in the LG can be resolved into stars, and their SFH can be derived (Skillman et al. 2014). Irrespective of their morphological type, all DGs shows signature of old stellar populations, although in different amount
(Weisz et al. 2014). This implies that DGs started to form stars at a sharply defined early epoch. The subsequent SFH was shaped by a variety of processes at work in the LG environment: ram pressure stripping, internal stellar evolution, harassment.  SFH
have been derived for many DG in the LG so far, and indeed present a large variety of shapes. Beside modelling SFH,
these processes also changed, dramatically in some cases, the DGs structure.
We described these processes in details and show example of them in the actual LG. 
Overall, the LG turns out to be a vibrant environment, where several processes conspire to produce structural modifications in dwarf galaxies, and shape their SFH. 
As already emphasised in the past, gas removal is the trigger of any transformation, and this occurs differently from dwarf to dwarf, depending on their mass, orbit, and properties of the medium they are travelling in.
A closer look at each individual DG will allow us in the future to understand better how they formed and evolved. Neutral and excited gas
observations are crucial (e.g., ALFALFA, LITTLE THING) to map the gas  structure and dynamics in and around DGs. Lastly, deep photometry is needed to extend the SFH derivation to the lowest-mass DG, because of their profound cosmological importance.

\begin{acknowledgement}
I warmly thank the organisers for inviting me to this great conference, and the ESO office for Science in Chile for the generous financial support. I also thank A.  McConnachie and Y. Momany for fruitful discussions.
\end{acknowledgement}
\eject

%
%
%

\end{document}